\documentclass{article}

\usepackage{arxiv}

\usepackage[utf8]{inputenc}
\usepackage[T1]{fontenc}
\usepackage[english]{babel}
\newenvironment{abstractbox}[1]
  {\centerline{\large\bfseries\scshape #1}\begin{quote}}
  {\end{quote}}
\usepackage{hyperref}
\usepackage{url}
\usepackage{booktabs}
\usepackage{amsmath}
\usepackage{amssymb}
\usepackage{microtype}
\usepackage{graphicx}
\usepackage{xcolor}
\usepackage{placeins}
\usepackage{multirow}
\usepackage{array}
\usepackage[round,semicolon]{natbib}
\usepackage{doi}

\newcommand{\figura}[2][\linewidth]{%
  \IfFileExists{#2}{\includegraphics[width=#1]{#2}}{%
    \fbox{\parbox[c][0.22\textheight][c]{0.92\linewidth}{\centering\small
      Missing figure:\\[2pt]\texttt{\detokenize{#2}}}}}}

\newcommand{\SarAIC}{464.24}  \newcommand{\SarBIC}{474.58}
\newcommand{\SarAICnat}{452.48}
\newcommand{\SarRMSE}{2.481}  \newcommand{\SarMAE}{1.557}
\newcommand{\SarRdos}{0.708}  \newcommand{\SarLB}{0.977}
\newcommand{\SarKurt}{16.64}  \newcommand{\SarSkew}{-2.64}
\newcommand{\SarK}{4}
\newcommand{\AsvAIC}{504.37}  \newcommand{\AsvBIC}{525.05}
\newcommand{\AsvRMSE}{2.923}  \newcommand{\AsvMAE}{1.831}
\newcommand{\AsvRdos}{0.594}  \newcommand{\AsvLB}{0.097}
\newcommand{\AsvK}{8}         \newcommand{\TauEst}{3.06}
\newcommand{\GpAIC}{508.15}   \newcommand{\GpBIC}{531.41}
\newcommand{\GpRMSE}{2.950}   \newcommand{\GpMAE}{1.814}
\newcommand{\GpRdos}{0.587}   \newcommand{\GpLB}{0.091}
\newcommand{\GpK}{9}          \newcommand{\GpCob}{90.0}
\newcommand{\PoiAIC}{509.06}  \newcommand{\PoiBIC}{534.91}
\newcommand{\PoiRMSE}{2.934}  \newcommand{\PoiMAE}{1.857}
\newcommand{\PoiRdos}{0.591}  \newcommand{\PoiLB}{0.133}
\newcommand{\PoiK}{10}        \newcommand{\TauJ}{1.25}
\newcommand{\PoiMu}{0.0513}   \newcommand{\PoiLamJ}{2.180}
\newcommand{\PoiKJ}{3.730}
\newcommand{\OosPoi}{4.037}   \newcommand{\OosAsv}{3.963}
\newcommand{\Nobs}{99}        \newcommand{\Neval}{98}
\newcommand{\MediaSerie}{3.39}\newcommand{\CurtSerie}{5.36}

\title{Shock covariates and continuous memory:\\
Volterra versus SARIMA for Ecuador's GDP}

\date{October 2026}

\author{%
  Luis Diego Campuzano Almeida%
  \thanks{This paper is derived from the author's thesis for the Master's
  in Applied Statistics at Escuela Superior Polit\'ecnica del Litoral
  (ESPOL).} \\
  Facultad de Ciencias Naturales y Matem\'aticas\\
  Escuela Superior Polit\'ecnica del Litoral (ESPOL)\\
  Guayaquil, Ecuador\\
  \texttt{luidcamp@espol.edu.ec}
}

\renewcommand{\shorttitle}{Shock covariates: Volterra versus SARIMA}
\renewcommand{\headeright}{Preprint}
\renewcommand{\undertitle}{Preprint}

\hypersetup{
  colorlinks=true, linkcolor=blue!50!black, citecolor=blue!50!black, urlcolor=blue!50!black,
  pdftitle={Shock covariates and continuous memory: Volterra versus SARIMA for Ecuador's GDP},
  pdfsubject={econ.EM; stat.AP; stat.ME},
  pdfauthor={Luis Diego Campuzano Almeida},
  pdfkeywords={Volterra integro-differential equations, SARIMA, SARIMAX, Gaussian processes, compound Poisson process, shock covariates, GDP, Ecuador},
}

\begin{document}
\maketitle

% ====================================================================
\begin{abstractbox}{Abstract}
We compare an augmented structural Volterra (ASV) integro-differential
framework, with a Gamma memory kernel and two stochastic extensions
(a Gaussian process and Poisson jumps), against a SARIMA$(1,0,0)(1,0,1)_4$
for the year-on-year growth of Ecuador's quarterly real GDP (2001Q1 to
2025Q3; \Nobs{} observations), and we extend both approaches by linking
each documented shock to the covariate that originates it: the COVID-19
stringency index, the oil price and a global financial indicator. In the
univariate setting, the SARIMA dominates the point metrics (AIC \SarAIC{}
versus \AsvAIC); the ASV models place the peak of shock propagation
between 3 and 4 quarters, produce 90\,\% intervals that cover 18 of 20
out-of-sample observations, and do not treat the 2008--2009 crisis as an
anomaly. The shock covariates change the diagnosis: with stringency and oil,
the SARIMAX AIC falls to 402.86, residual kurtosis drops from 16.6 to 1.3
and normality is no longer rejected, which the jump mechanism did not
achieve; the financial indicator is not significant, consistent with the
absence of an anomaly in 2008--2009. The covariates reduce the SARIMA error
at every horizon of a rolling origin; however, against the data published
for 2025Q4 to 2026Q2 and with the observed covariates, the univariate ASV
forecasts better (MAE 0.78 pp) than the SARIMAX (3.30) and the ASV-X
(1.30). Covariates explain past shocks; continuous memory sustains
multi-step forecasting better.
\end{abstractbox}

\def\keywordname{{\bfseries\emph{Keywords}}}
\keywords{Volterra integro-differential equation \and SARIMA \and Gaussian
process \and Compound Poisson process \and Long memory \and Ecuador GDP
\and Shock covariates \and SARIMAX}


% ====================================================================
\section{Introduction}
\label{sec:intro}
% ====================================================================

Modelling economic growth is a central topic in applied macroeconomics,
especially for developing economies exposed to external shocks, structural
change and persistent volatility \citep{asteriou2021,vanderploeg2009}. The
Ecuadorian case adds an institutional complication: officially dollarised
since 2000, the country has neither its own monetary policy nor a lender of
last resort \citep{quispe2006}, so its growth path depends to a large
extent on structural inertia, fiscal policy and the transmission of
commodity price fluctuations.

SARIMA models have become a standard tool for macroeconomic series because
of their flexibility and solid empirical performance
\citep{box2015,hyndman2021}. However, they rely on finite, discrete lag
structures that impose a \emph{memory truncation}, which may be
insufficient when the dynamics exhibit deep persistence or hysteresis that
fades gradually \citep{diebold1989,baillie1996}. In response, a growing
body of research has explored non-local, integro-differential and
fractional specifications that weight the past continuously through a
memory kernel \citep{tarasov2018,tarasova2018}. This feature is
particularly relevant when shocks propagate slowly through investment,
employment and trade \citep{granger1980}, as one would expect in an economy
without a lender of last resort.

This paper starts from a deterministic augmented structural Volterra (ASV)
model that combines sinusoidal seasonal forcing, an annual delay of the
delay-differential-equation type and a non-monotonic Gamma memory kernel,
and extends it in two stochastic directions: (i)~a memory kernel governed
by a Gaussian process (ASV-GP), which propagates uncertainty about the
memory weights to the forecast \citep{rasmussen2006}; and (ii)~a compound
Poisson process with persistent effect (ASV-Poisson), which detects and
characterises large discrete shocks \citep{merton1976}. The four models are
compared on a harmonised likelihood basis, a condition that, as we show, is
indispensable for the AIC to be comparable across model families.

The research question is twofold: whether the intervals produced by the
ASV-GP are empirically calibrated out of sample, and whether the
ASV-Poisson identifies, without systematic omissions or false alarms, the
documented break episodes in Ecuadorian GDP (2008--2009, 2015--2016 and
2020). Three hypotheses are tested: \textbf{H1}, the SARIMA outperforms the
ASV models in in-sample accuracy and statistical efficiency; \textbf{H2},
the 90\,\% ASV-GP intervals have empirical coverage close to the nominal
level; and \textbf{H3}, the ASV-Poisson identifies the three documented
episodes (H3a) and improves residual normality relative to the base ASV
(H3b). As an extension beyond these hypotheses, we examine whether linking
each shock to an observable covariate that originates it ---instead of
treating it as a random jump--- better explains the break episodes and
improves conditional forecasting.

The contributions are four. First, a systematic, harmonised comparison
between an integro-differential framework with stochastic components and
the SARIMA, which remains scarce in the applied literature. Second,
evidence of out-of-sample calibration of intervals that incorporate
uncertainty about the memory structure, robust at 90\,\% to the strict
re-estimation of all parameters. Third, an auditable catalogue of breaks,
with an exhaustive threshold-sensitivity analysis, whose most informative
result is negative: the 2008--2009 crisis is not an anomaly relative to the
internal dynamics of Ecuadorian GDP. Fourth, an extension with shock
covariates ---one per episode--- that explains the 2020 collapse with the
stringency index and the 2015--2016 slump with the oil price, removes the
residual leptokurtosis, and is evaluated against the data published after
the sample. Results contrary to the hypotheses are also reported without
attenuation: H1 and H2 are confirmed, and H3 is rejected in both of its
components.

The rest of the paper is organised as follows. Section~\ref{sec:lit}
reviews the literature; Section~\ref{sec:datos} describes the data;
Section~\ref{sec:metodo} presents the models and the evaluation protocol;
Section~\ref{sec:resultados} reports the univariate results;
Section~\ref{sec:covariables} presents the extension with shock covariates;
Section~\ref{sec:discusion} discusses both; and
Section~\ref{sec:conclusiones} concludes.

% ====================================================================
\section{Related literature}
\label{sec:lit}
% ====================================================================

\paragraph{Linear seasonal models.} ARIMA and SARIMA models are the dominant
empirical framework for GDP growth \citep{box2015}, and they remain a
benchmark that is hard to beat in forecasting economic series
\citep{hyndman2021,petropoulos2022}, partly because their AR and MA
components have a direct interpretation in terms of inertia and shock
propagation. Their main limitation is that the influence of the past is
truncated at a finite number of lags \citep{baillie1996}.

\paragraph{Persistence and long memory.} The persistence of macroeconomic
series is widely documented: \citet{diebold1989} find long memory in
aggregate output, and \citet{granger1980} shows that aggregation can
generate it. Processes with hyperbolically decaying autocorrelations have
been systematised by \citet{baillie1996}, \citet{palma2007} and
\citet{beran2013}, and \citet{gilalana1997} test them in macroeconomic
series through fractional integration tests. In commodity-dependent
economies, moreover, growth is more volatile \citep{vanderploeg2009}.
Estimation of the memory order, however, is imprecise in short samples
\citep{beran2013}, a relevant restriction with \Nobs{} observations.

\paragraph{Integro-differential models and non-monotonic kernels.}
\citet{tarasov2018} formalise the concept of dynamic memory in economics
through memory functions that weight the history of the process, and
\citet{tarasova2018} apply it to macroeconomic models with fractional
calculus. This paper adopts a Gamma-type kernel, which allows a delayed
impact with an interior peak instead of the monotonic decay of fractional
kernels. That shape connects with the phenomenon of \emph{transmission
lags}: the maximum effect of a policy shock is not instantaneous
\citep{havranek2013}. The representation is relevant in Ecuador, where
dollarisation removes monetary adjustment as a shock absorber
\citep{quispe2006}.

\paragraph{Pandemic shocks and model selection.} Studies of the COVID-19
shock document broad and heterogeneous effects \citep{brodeur2021}, but
they rarely evaluate the calibration of uncertainty or the accuracy with
which known shocks are detected. \citet{petropoulos2022} stress the
importance of model uncertainty in forecasting, and \citet{hendry2014}
argue that model selection should not rest on a fit criterion alone, but
also on the evaluation of the model's properties against the data
generating process. This paper puts that argument into practice.

\paragraph{Exogenous shocks and covariates.} The intervention analysis of
\citet{boxtiao1975} formalises the inclusion of exogenous events in ARIMA
models through explanatory variables or transfer functions. For
oil-exporting economies, oil price shocks are a documented channel of
recession \citep{hamilton1983}, and for the pandemic, the stringency index
of \citet{hale2021} offers a comparable measure of the intensity of
containment policies, closer to the economic channel than counts of
infections or deaths.

% ====================================================================
\section{Data and stylised facts}
\label{sec:datos}
% ====================================================================

The variable is the year-on-year growth rate of Ecuador's quarterly real
GDP, taken from the National Accounts Statistics of the Central Bank of
Ecuador \citep{bce2025}, compiled under the SNA 2008. The sample covers
2001Q1 to 2025Q3 ($N=\Nobs$); it starts in 2001 because that is the first
full year of the dollarised regime, which guarantees monetary homogeneity.
Table~\ref{tab:descriptivos} summarises the distribution for the full
sample and by volatility regime.

\begin{table}[htbp]
\centering
\caption{Descriptive statistics of the year-on-year growth of Ecuador's
quarterly real GDP (\%).}
\label{tab:descriptivos}
\small
\begin{tabular}{lcccccc}
\toprule
 & Obs. & Mean & Median & Std.\ dev. & Min. & Max. \\
\midrule
Total 2001--2025                          & \Nobs & \MediaSerie & 4.10 & 4.60 & $-18.0$ & 19.3 \\
P1: 2001--2008 (dollarisation and boom)   & 32 & 4.64 & --- & 1.99 & 1.2     & 7.8 \\
P2: 2009--2019 (adjustment and shocks)    & 44 & 3.43 & --- & 3.49 & $-3.4$  & 9.5 \\
P3: 2020--2025 (pandemic and recovery)    & 23 & 1.58 & --- & 7.70 & $-18.0$ & 19.3 \\
\bottomrule
\end{tabular}

\smallskip
{\footnotesize Note: full series with skewness $-0.93$, excess kurtosis
\CurtSerie{} and Shapiro-Wilk $W=0.9186$ ($p<0.001$); the minimum
corresponds to 2020Q2 and the maximum to 2021Q2. Source: \citet{bce2025};
own elaboration.}
\end{table}

\begin{figure}[htbp]
\centering
\figura{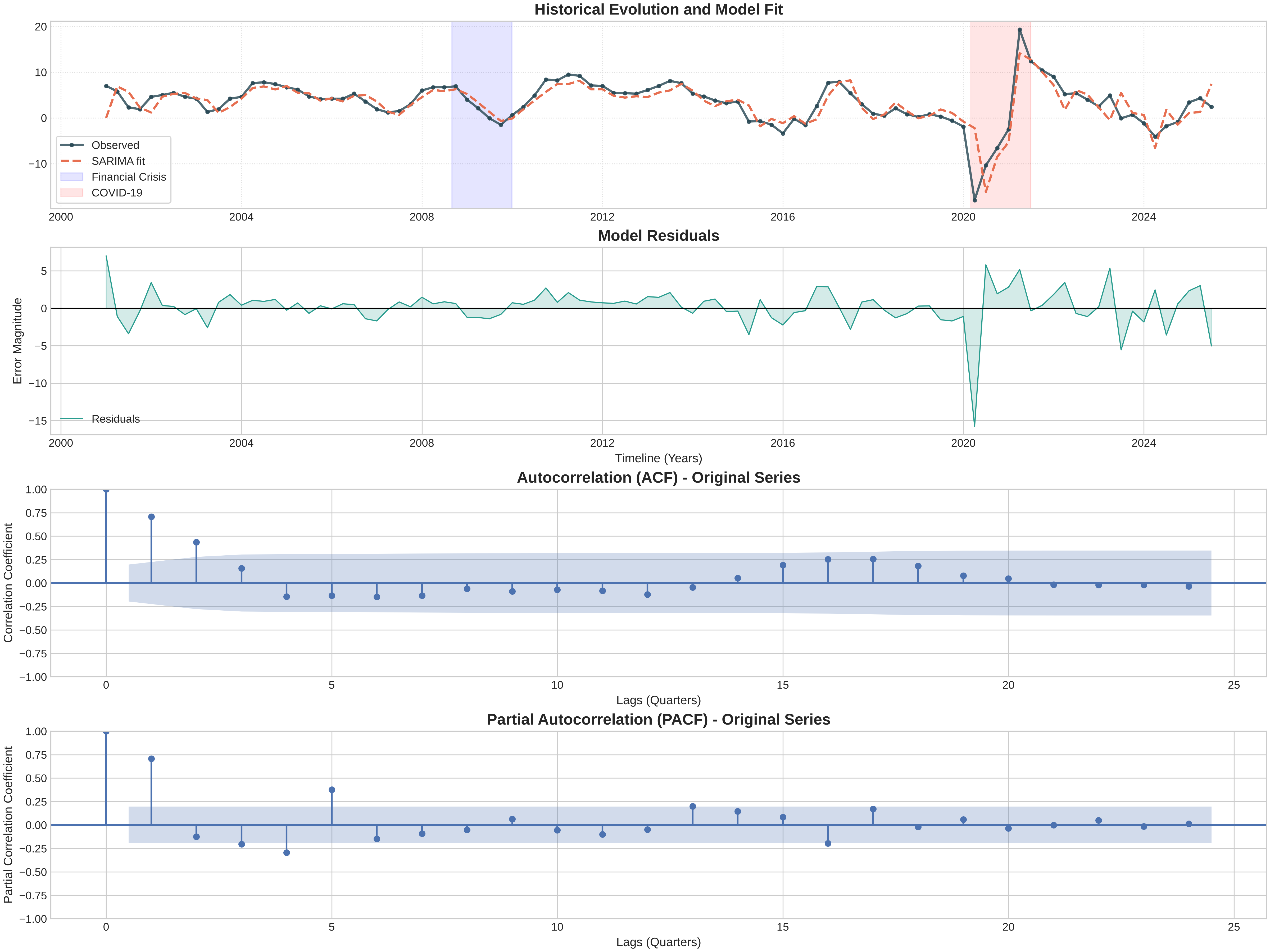}
\caption{Year-on-year growth of Ecuador's quarterly real GDP (2001Q1 to
2025Q3), baseline SARIMA fit, fit residuals, and ACF and PACF of the
original series.}
\label{fig:serie}
\end{figure}

Three features shape the modelling. A mean below the median and negative
skewness indicate contractions more extreme than expansions, consistent
with a commodity-exporting economy. Excess kurtosis (\CurtSerie) reveals
heavy tails: excluding 2020Q2 and 2021Q2 it falls to 1.13, which suggests
discrete events rather than a continuous tail and motivates the jump
component. Finally, the standard deviation of regime P3 is 3.9 times that
of P1, a sign of heteroskedasticity across regimes.
Figure~\ref{fig:serie} shows the series and the baseline linear
fit.

\paragraph{Stationarity.} The ADF and KPSS tests agree that the series is
$I(0)$ (Table~\ref{tab:estacionariedad}); the verdict is replicated on the
first difference, which rules out ambiguity about the order of
integration. This is consistent with the use of year-on-year rates, which
act as a seasonal difference of the log level. We therefore set $d=D=0$.

\begin{table}[htbp]
\centering
\caption{Unit-root and stationarity tests.}
\label{tab:estacionariedad}
\small
\begin{tabular}{llcccc}
\toprule
Test & $H_0$ & Statistic & $p$-value & 5\,\% CV & Decision \\
\midrule
ADF (levels)            & Unit root     & $-3.1407$ & $0.0237$    & $-2.893$ & Reject \\
KPSS (levels)           & Stationarity  & $0.3149$  & $\geq 0.10$ & $0.463$  & Do not reject \\
ADF (first difference)  & Unit root     & $-5.2056$ & $<0.001$    & $-2.896$ & Reject \\
KPSS (first difference) & Stationarity  & $0.0178$  & $\geq 0.10$ & $0.463$  & Do not reject \\
\bottomrule
\end{tabular}

\smallskip
{\footnotesize Note: ADF lags selected by AIC (4 in levels, 12 in
differences); KPSS with 5 and 0 lags. Source: own elaboration.}
\end{table}

\paragraph{Autocorrelation, seasonality and spectrum.} The ACF is
significant at lags 1 and 2 ($|r_k|>0.197$) and then decays gradually; the
PACF shows a dominant spike at lag 1 and secondary spikes at lags 3 and 4
(negative) and 5 (positive). An ANOVA on the quarterly means detects no
differences ($F=0.0406$, $p=0.989$), and the additive seasonal component
ranges only between $-0.10$ and $0.09$ pp: the remaining seasonality is
stochastic, not deterministic. The periodogram concentrates its power at
$f\approx0.061$ cycles per quarter (a cycle of about four years), whereas
the seasonal frequency $f=0.25$ accounts for only 0.17\,\% of the power in
the band $[0.23,0.27]$.

\paragraph{Long memory.} The rescaled-range estimator gives $\hat H=0.716$
($d=\hat H-0.5=0.216$), compatible with long-range dependence, although the
R/S estimator has a known positive bias in short samples. This motivates
modelling persistence with a parametric memory kernel rather than with
fractional differencing, whose explicit variant is tested in
Section~\ref{sec:fraccionario}.

\begin{figure}[htbp]
\centering
\figura{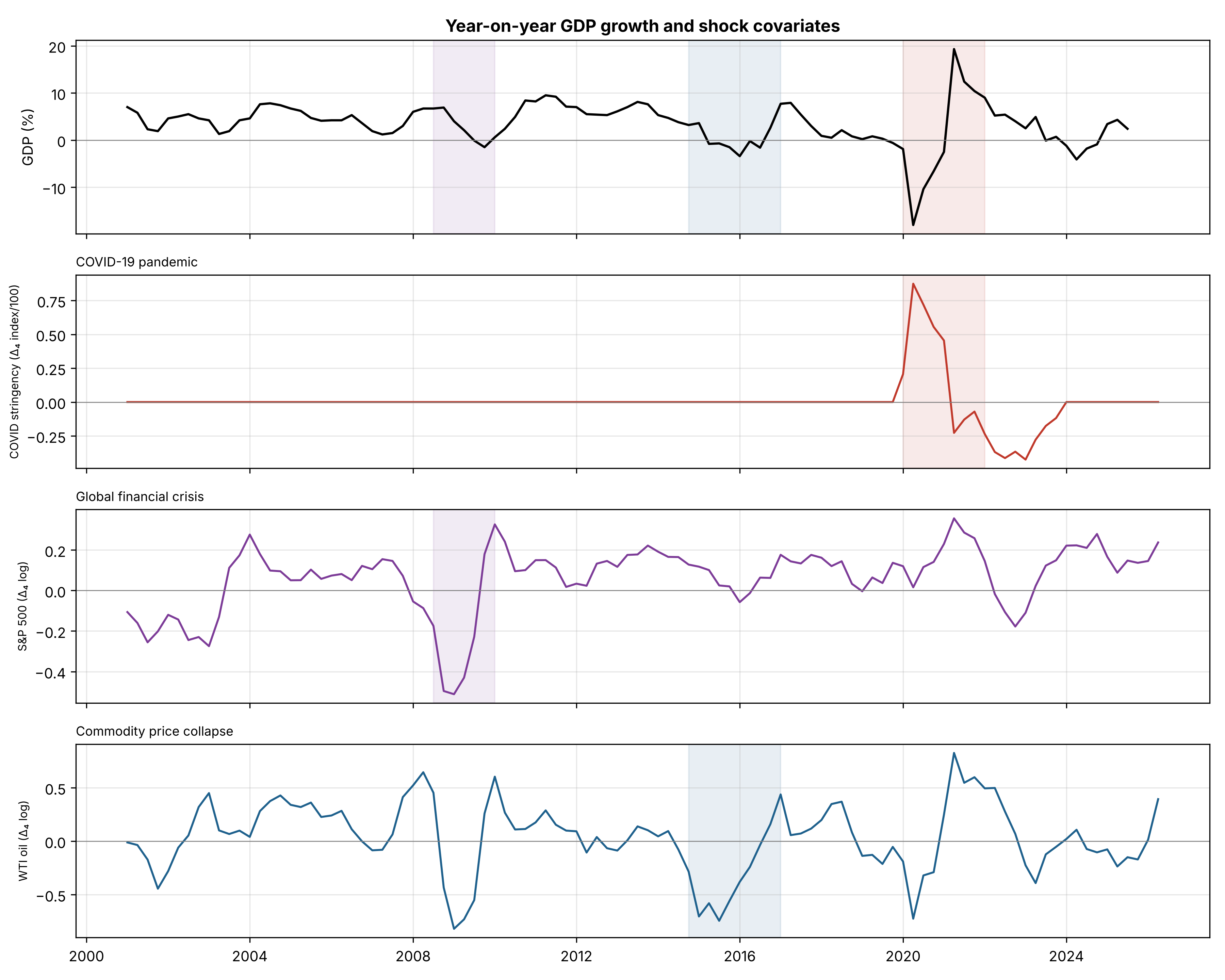}
\caption{Year-on-year GDP growth and shock covariates. Shaded areas mark the
window of each shock, used for descriptive purposes only.}
\label{fig:covariables}
\end{figure}

\paragraph{Shock covariates.} Each documented break episode is linked to
the covariate that originates it (Table~\ref{tab:choques},
Figure~\ref{fig:covariables}). All covariates are aggregated to quarterly
frequency and brought to the year-on-year scale of GDP. The stringency
index of \citet{hale2021} covers 2020Q1 to 2022Q4; before 2020 it equals
zero and, after 2022, it is assumed to be zero, given the end of
containment measures. The financial covariate is included to test whether
the 2008--2009 crisis, which the jump detector did not flag, responds to an
observable external factor.

\begin{table}[htbp]
\centering
\caption{Shock--covariate register.}
\label{tab:choques}
\small
\begin{tabular}{p{4.6cm}llc}
\toprule
Shock & Covariate & Source & Transformation \\
\midrule
Global financial crisis (2008--2009)   & S\&P 500            & \citet{shiller2015} & $\ln x_t-\ln x_{t-4}$ \\
Commodity price slump (2015--2016)     & WTI oil price       & \citet{eia2026}     & $\ln x_t-\ln x_{t-4}$ \\
COVID-19 pandemic (2020--2021)         & Stringency index    & \citet{hale2021}    & $(S_t-S_{t-4})/100$ \\
\bottomrule
\end{tabular}

\smallskip
{\footnotesize Note: quarterly averages of daily or monthly data; data
available up to 2026Q3 (WTI and S\&P 500) and 2022Q4 (stringency). Source:
own elaboration.}
\end{table}

% ====================================================================
\section{Methodology}
\label{sec:metodo}
% ====================================================================

\subsection{Reference model: SARIMA}

The SARIMA$(p,d,q)(P,D,Q)_4$ is specified as
\begin{equation}
\Phi_P(L^4)\,\phi_p(L)\,(1-L)^d(1-L^4)^D\,y_t=\Theta_Q(L^4)\,\theta_q(L)\,\varepsilon_t,
\label{eq:sarima}
\end{equation}
where $L$ is the lag operator and $\varepsilon_t$ is white noise. Five
candidate specifications are evaluated: two suggested by ACF/PACF
identification (M1, M2) and three with seasonal differencing or orders
outside the suggested range (M3--M5), included as a robustness check.
Estimation is by exact maximum likelihood through the Kalman filter
(\texttt{statsmodels}, \texttt{SARIMAX}); details are given in
Appendix~\ref{app:sarima}.

\subsection{Augmented structural Volterra (ASV) framework}

Let $y(t)$ be the growth rate. The series is centred,
$y_c(t)=y(t)-\bar\mu$ with $\bar\mu=\MediaSerie\,\%$, and the mean is
restored in each prediction. The general equation, on the centred series,
is
\begin{equation}
y_c(t)=(1+\alpha)\,y_c(t-1)+\gamma\,y_c(t-4)
+\beta\!\!\sum_{\tau\notin\{1,4\}}\!\!K(\tau)\,y_c(t-\tau)
+A\cos\!\Big(\frac{2\pi t}{4}+\varphi\Big)+J(t)+\varepsilon(t),
\label{eq:asv}
\end{equation}
where $\alpha$ captures inertia, $\gamma$ the annual delay, $K(\tau)$ the
memory kernel (with a window of $L=24$ lags for fitting and evaluation, and
of 12 lags inside the objective function during optimisation), $A$ and
$\varphi$ the amplitude and phase of the seasonal forcing, and $J(t)$ the
jump component. The sum excludes $\tau\in\{1,4\}$ to ensure the
identifiability of $\alpha$ and $\gamma$. Parameters are estimated by
minimising the squared error with global Differential Evolution followed
by L-BFGS-B refinement.

\paragraph{Deterministic ASV.} The kernel is a discrete Gamma,
\begin{equation}
K(\tau)=\tau^{k-1}e^{-\lambda\tau},\qquad
\tau^*=\frac{k-1}{\lambda},
\label{eq:gamma}
\end{equation}
where $\tau^*$ is the lag of maximum impact (first-order condition) and
$J(t)\equiv0$. Including the intercept, the model has $k=\AsvK$
parameters.

\paragraph{ASV-GP.} The weights are treated as a random function,
$K(\tau)\sim\mathcal{GP}\big(m(\tau),\,c(\tau,\tau')\big)$
\citep{rasmussen2006}, with two candidate covariances:
\begin{equation}
c_{\mathrm{RBF}}=\sigma_f^2\exp\!\Big(-\frac{r^2}{2\ell^2}\Big),\qquad
c_{\mathrm{M52}}=\sigma_f^2\Big(1+\frac{\sqrt5\,r}{\ell}+\frac{5r^2}{3\ell^2}\Big)e^{-\sqrt5 r/\ell},
\qquad r=|\tau-\tau'|.
\label{eq:gp}
\end{equation}
Because the weights are latent inside recursion~\eqref{eq:asv}, there is no
closed-form marginal likelihood. We adopt an empirical-Bayes approximation:
a fixed Gamma-type guide function, $g(\tau)=\tau^{1.5}e^{-0.5\tau}$
normalised (peaking at $\tau=3$), acts as pseudo-observations, and the GP
posterior mean over lags 2 to 16 is conditioned on it, with
hyperparameters $(\ell,\sigma_f,\sigma_n)$ fixed at their optimum; the
objective function adds to the squared error a regularisation term equal to
$0.005$ times the GP log marginal likelihood on the guide function.
Intervals are obtained from 300 trajectories that perturb the posterior
mean with noise correlated according to the GP covariance, scaled by a
factor of 0.3. The model has $k=\GpK$ parameters (five structural, three
hyperparameters and the intercept).

\paragraph{ASV-Poisson.} Discrete shocks arrive according to a compound
Poisson process \citep{merton1976} and persist according to a decay
kernel:
\begin{equation}
J(t)=\sum_{s<t}J_s\,K_J(t-s),\qquad K_J(\tau)=\tau^{k_J-1}e^{-\lambda_J\tau},
\qquad \tau^*_J=\frac{k_J-1}{\lambda_J}.
\label{eq:poisson}
\end{equation}
Shock periods are the residuals of the base ASV that exceed $\pm2\sigma$,
with the robust scale $\sigma_{\mathrm{MAD}}=1.4826\cdot\mathrm{MAD}$: the
classical standard deviation is contaminated by the very extreme residuals
one wants to detect. Threshold sensitivity is reported in full. The model
has $k=\PoiK$ parameters.

\subsection{Comparability of information criteria}
\label{sec:aic}

The four models are evaluated on the same set of $N_{\mathrm{eval}}=\Neval$
observations and with the same full Gaussian likelihood:
\begin{equation}
\mathrm{AIC}=n\ln\!\Big(\frac{\mathrm{RSS}}{n}\Big)+n\,(1+\ln2\pi)+2k,
\qquad n=N_{\mathrm{eval}}.
\label{eq:aic}
\end{equation}
Harmonisation is not a technical detail. The native AIC of
\texttt{statsmodels} includes the constant $n(1+\ln2\pi)\approx278.1$,
whereas a naive calculation from the RSS of the ASV models omits it;
comparing both figures directly produces a spurious advantage of about 278
units in favour of the ASV framework. All criteria in this paper are
computed with~\eqref{eq:aic}; the native SARIMA AIC (\SarAICnat) is not
comparable and is reported only for reference.

\subsection{Out-of-sample validation}

To avoid data snooping, we use an 80/20 temporal split with
$t_{\mathrm{split}}=\lfloor0.8\times\Nobs\rfloor=79$: shocks are detected,
and $\lambda_J$, $k_J$ and $\hat\sigma_\varepsilon$ are estimated, only
with $t<79$; the segment $t\geq79$ (20 observations) is reserved to
evaluate the out-of-sample RMSE and the empirical coverage of the ASV-GP
intervals against the nominal level. In this baseline protocol, the
structural and GP parameters are estimated with the full sample;
Section~\ref{sec:robustez} repeats the evaluation re-estimating all
parameters with $t<79$ only.

\subsection{Extension with shock covariates}
\label{sec:metodo_cov}

Let $x_{m,t}$ be the covariate associated with shock $m$. Since
$y_t\approx100\,(\ln Y_t-\ln Y_{t-4})$, a covariate that shifts the
\emph{level} of GDP must enter through its year-on-year change. For
stringency, if $\ln Y_t$ falls in proportion to the index
$S_t\in[0,100]$, the effect on growth is proportional to
$x_t=(S_t-S_{t-4})/100$: positive when measures are imposed and negative
when they are lifted, which explains both the 2020 collapse and the 2021
base effect. Death counts, in levels, do not follow this logic: their
maximum (2021Q3) coincides with growth above 10\,\%.

The SARIMAX specification is a regression with SARIMA errors of the same
order as the reference model,
\begin{equation}
y_t=\sum_m \beta_m\,x_{m,t}+u_t,\qquad
\phi(L)\,\Phi(L^4)\,u_t=\Theta(L^4)\,\varepsilon_t,
\label{eq:sarimax}
\end{equation}
and the ASV-X adds to equation~\eqref{eq:asv} a term
$\sum_m \eta_m\,x_{m,t}$ in place of the jump component $J(t)$, so that the
shock is attributed to a measured variable rather than to a random event.
In an exploratory test, not included in the replication code, a variant
with a Gamma lag kernel of its own for each covariate converged to the
contemporaneous effect (the decay reaches its search bound and the AIC
worsens), so the contemporaneous form is adopted. The final specification
is chosen by harmonised AIC and evaluated with the 80/20 split, with an
expanding-window rolling origin (horizons 1 to 4), and with the GDP data
published after the sample.

When forecasting, the \emph{observed} values of each covariate over the
horizon are used: the forecast is conditional on the realisation of the
factor that originates each shock. Beyond the last available data point, a
scenario in levels is adopted: zero stringency and oil held constant at its
last quarterly value.

% ====================================================================
\section{Results}
\label{sec:resultados}
% ====================================================================

\subsection{Reference SARIMA}

Table~\ref{tab:sarima} shows that M1, SARIMA$(1,0,0)(1,0,1)_4$, minimises
the harmonised AIC with an advantage of $\Delta\mathrm{AIC}=29.94$ over the
second candidate, well above conventional thresholds \citep{burnham2002}.
The candidates with seasonal differencing (M3--M5) are heavily penalised,
which validates ex post the tests of Section~\ref{sec:datos}.

\begin{table}[htbp]
\centering
\caption{SARIMA selection (harmonised AIC, $N_{\mathrm{eval}}=\Neval$).}
\label{tab:sarima}
\small
\begin{tabular}{llcccc}
\toprule
ID & Specification & $k$ & AIC (harm.) & RMSE (pp) & $R^2$ \\
\midrule
M1 & SARIMA$(1,0,0)(1,0,1)_4$ & 4 & \textbf{\SarAIC} & \textbf{\SarRMSE} & \textbf{\SarRdos} \\
M2 & SARIMA$(1,0,1)(1,0,0)_4$ & 4 & 494.18 & 2.891 & 0.603 \\
M3 & SARIMA$(1,0,1)(1,1,1)_4$ & 5 & 497.02 & 2.903 & 0.600 \\
M4 & SARIMA$(2,0,2)(1,1,1)_4$ & 7 & 498.98 & 2.873 & 0.608 \\
M5 & SARIMA$(0,0,1)(0,1,1)_4$ & 3 & 536.44 & 3.623 & 0.376 \\
\bottomrule
\end{tabular}

\smallskip
{\footnotesize Note: $k$ = estimated parameters, including the variance
$\sigma^2_\varepsilon$ (Appendix~\ref{app:sarima}). Source: own
elaboration.}
\end{table}

The estimates (Table~\ref{tab:parametros}, panel A) are more nuanced than
the selection suggests. Only $\phi_1=0.984$ is individually significant,
and it lies close to the unit root: without a constant term, the
autoregressive component also absorbs the mean level of the series. The
SAR(1) is not significant ($p=0.254$) and the SMA(1) converges to the
non-invertibility boundary ($\Theta_1=-1$), a \emph{pile-up} phenomenon
that invalidates individual Wald tests; the seasonal structure must
therefore be judged by its joint contribution to the likelihood. The
residuals pass the white-noise test (Ljung-Box $p=\SarLB$), but strongly
reject normality: excess kurtosis \SarKurt, skewness $\SarSkew$ and
Shapiro-Wilk $p<0.001$.

\subsection{Deterministic ASV: transmission scale}

Table~\ref{tab:parametros} (panel B) reports the structural parameters. The
lag of maximum impact of the kernel is
\begin{equation}
\tau^*=\frac{k-1}{\lambda}=\frac{4.0000-1}{0.9805}=\TauEst\ \text{quarters}.
\end{equation}
The maximum impact of a shock on growth thus appears about three quarters
after the event, and $k>1$ confirms the non-monotonic profile. The
economic reading is consistent with dollarisation: without its own
monetary policy, the real sector absorbs shocks through adjustments in
relative prices, credit and trade, which take longer. This reading should
be taken with caution: the point value of $\tau^*$ depends on the search
bound for $k$, and memory is identified only when the sample includes the
pandemic cycle (Section~\ref{sec:robustez}). Two further features are
worth noting. The effective coefficient on the first lag
($1+\alpha=0.696$) is much lower than the SARIMA $\phi_1$, because the ASV
splits persistence among inertia, annual delay, memory and an explicit
intercept. And $\gamma<0$ acts as an opposite-sign correction on the level
of a year earlier, analogous to the seasonal SMA(1).

\begin{table}[htbp]
\centering
\caption{Estimated parameters of the optimal SARIMA and the deterministic
ASV.}
\label{tab:parametros}
\small
\begin{tabular}{llccl}
\toprule
\multicolumn{5}{l}{\textit{Panel A. SARIMA$(1,0,0)(1,0,1)_4$ (exact ML; OPG standard errors)}} \\
\midrule
Parameter & & Estimate & SE & $p$ \\
\midrule
AR(1) & $\phi_1$     & $0.9836$  & $0.0109$ & $<0.001$ \\
SAR(1) & $\Phi_1$    & $-0.0997$ & $0.0874$ & $0.254$ \\
SMA(1) & $\Theta_1$  & $-1.0000$ & $264.19^{\dagger}$ & $0.997^{\dagger}$ \\
Variance & $\hat\sigma^2_\varepsilon$ & $5.780$ & --- & --- \\
\midrule
\multicolumn{5}{l}{\textit{Panel B. Deterministic ASV (Differential Evolution + L-BFGS-B)}} \\
\midrule
Parameter & Description & Estimate & & Interpretation \\
\midrule
$\alpha$    & Inertia             & $-0.3041$ & & $1+\alpha=0.696$ on $y(t-1)$ \\
$\gamma$    & Annual delay        & $-0.4427$ & & Negative seasonal correction \\
$\beta$     & Memory scale        & $0.0766$  & & Weight of continuous memory \\
$\lambda$   & Decay               & $0.9805$  & & Dissipation of the effect \\
$k$         & Gamma shape         & $4.0000^{\ddagger}$  & & Hump-shaped profile \\
$A$         & Seasonal amplitude  & $0.2659$  & & $\pm0.27$ pp \\
$\varphi$   & Seasonal phase      & $-0.4624$ & & rad \\
$\bar\mu$   & Intercept           & \MediaSerie & & Additional degree of freedom \\
\bottomrule
\end{tabular}

\smallskip
{\footnotesize Note: $^{\dagger}$~$\Theta_1$ at the non-invertibility
boundary; SE and Wald tests are not valid. $^{\ddagger}$~Upper bound of the
search space; see Section~\ref{sec:robustez}. Source: own elaboration.}
\end{table}

\subsection{ASV-GP: kernel selection and calibration}
\label{sec:gp}

\begin{figure}[htbp]
\centering
\figura{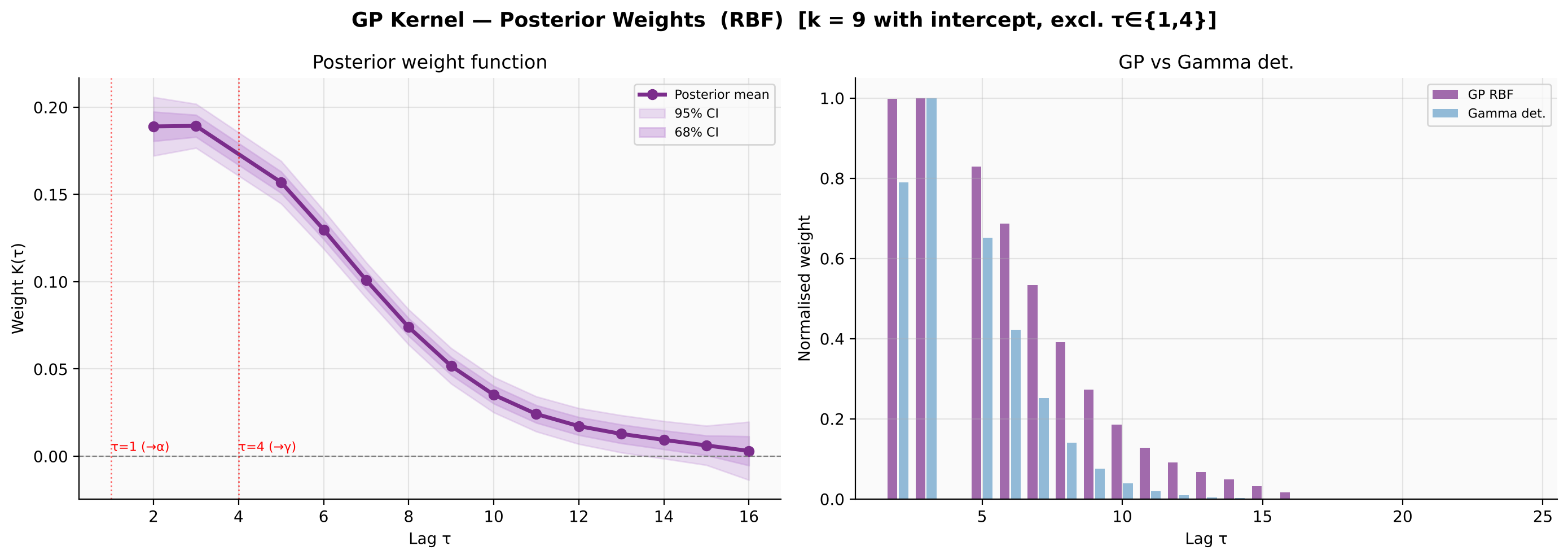}
\caption{Memory-kernel weights of the ASV-GP (RBF covariance): posterior
mean and credible bands.}
\label{fig:gpkernel}
\end{figure}

The RBF kernel minimises the harmonised AIC (Table~\ref{tab:gp}, panel A),
but the difference from the Matérn 5/2 is only 0.05 units: both are
empirically indistinguishable and the choice is a technical tie-break. The
length scale ($\ell=5.03$ quarters) indicates strongly correlated weights
across neighbouring lags, consistent with a memory that varies gradually.
The posterior mean concentrates the largest weight at lags 2 and 3, in line
with the peak of the guide function, and then decays more slowly than the
deterministic Gamma kernel, that is, with a heavier tail; relative
uncertainty grows at long lags, where the estimated weight is small
(Figure~\ref{fig:gpkernel}).

\begin{table}[htbp]
\centering
\caption{ASV-GP: covariance selection and out-of-sample calibration.}
\label{tab:gp}
\small
\begin{tabular}{lccccc}
\toprule
\multicolumn{6}{l}{\textit{Panel A. Covariance functions ($k=\GpK$)}} \\
\midrule
Kernel & RMSE (pp) & MAE (pp) & $R^2$ & AIC (harm.) & Status \\
\midrule
RBF        & 2.950 & 1.814 & 0.587 & \textbf{508.15} & Selected \\
Matérn 5/2 & 2.951 & 1.815 & 0.587 & 508.19 & Discarded \\
\midrule
\multicolumn{6}{l}{\textit{Panel B. Coverage of one-step predictive intervals ($t\geq79$, $n=20$)}} \\
\midrule
Interval & Nominal & Empirical & Difference & Exact binomial $p$ & \\
\midrule
50\,\% & 50.0\,\% & 50.0\,\% (10/20) & 0.0 pp & 1.000 & \\
90\,\% & 90.0\,\% & \GpCob\,\% (18/20) & 0.0 pp & 1.000 & \\
\bottomrule
\end{tabular}

\smallskip
{\footnotesize Note: RBF with $\ell=5.026$, $\sigma_f=0.489$,
$\sigma_n=0.050$. Intervals combine 300 draws from the GP posterior with
residual noise $\mathcal{N}(0,\hat\sigma_\varepsilon^2)$,
$\hat\sigma_\varepsilon=2.58$ pp estimated on the training segment only.
Source: own elaboration.}
\end{table}

In the baseline protocol, calibration on the test segment (panel B) is
exact at both levels and confirms H2. When all parameters are re-estimated
with the training segment only, the 90\,\% interval keeps its 18/20
coverage and the 50\,\% interval drops to 7/20
(Section~\ref{sec:robustez}). A complementary result locates the source of
uncertainty: purely structural intervals, which propagate only the
uncertainty of the weights without residual noise, cover just 30\,\%
(90\,\% interval) and 20\,\% (50\,\% interval); most of the one-step
predictive uncertainty comes from residual noise, not from the parametric
uncertainty of the kernel. This proportion depends, however, on the 0.3
factor that scales the weight perturbations (Section~4.2). Moreover, the
validation segment (2020Q4 to 2025Q3) includes the 2021Q2 rebound but not
the 2020Q2 collapse, which falls in the training segment; with only 20
observations, calibration cannot be extrapolated to scenarios with breaks
of a magnitude comparable to 2020.

Note that the ASV-GP has the worst RMSE and the worst $R^2$ of the four
models, although the best MAE among the three ASV models: making the kernel
more flexible does not improve the point fit, and its contribution is
strictly inferential.

\subsection{ASV-Poisson: shock catalogue}
\label{sec:poisson}

The classical standard deviation of the training residuals ($\sigma=2.59$
pp) is inflated by the pandemic residual and raises the threshold to
$\pm5.18$ pp, reachable only by the very event that contaminates it; the
robust scale ($\sigma_{\mathrm{MAD}}=1.60$ pp) places it at $\pm3.20$ pp.
Table~\ref{tab:poisson} (panel A) reports the full sensitivity analysis, so
that the choice of threshold is auditable rather than an \emph{ad hoc} fit
to the known episodes.

\begin{table}[htbp]
\centering
\caption{ASV-Poisson: threshold sensitivity and shock catalogue (base ASV
residuals, training segment).}
\label{tab:poisson}
\small
\begin{tabular}{lccp{7.6cm}}
\toprule
\multicolumn{4}{l}{\textit{Panel A. Threshold sensitivity}} \\
\midrule
Threshold & $\pm$pp & No. & Quarters detected \\
\midrule
$2.0\sigma$ classical & 5.18 & 1 & 2020Q2 \\
$1.5\sigma$ classical & 3.88 & 2 & 2015Q2, 2020Q2 \\
$\mathbf{2.0\sigma_{MAD}}$ \textbf{(active)} & \textbf{3.20} & \textbf{4} & \textbf{2002Q1, 2015Q2, 2020Q1, 2020Q2} \\
$1.5\sigma_{\mathrm{MAD}}$ & 2.40 & 11 & 2002Q1, 2003Q2, 2010Q4, 2015Q2, 2016Q1, 2016Q3, 2017Q1, 2017Q3, 2019Q1, 2020Q1, 2020Q2 \\
\midrule
\multicolumn{4}{l}{\textit{Panel B. Catalogue with the active threshold}} \\
\midrule
Quarter & Residual (pp) & Sign & Documented episode \\
\midrule
2002Q1 & $+3.44$  & $+$ & None (post-dollarisation investment boom) \\
2015Q2 & $-4.46$  & $-$ & Commodity price slump 2015--2016 \\
2020Q1 & $-3.27$  & $-$ & COVID-19 pandemic (onset) \\
2020Q2 & $-17.85$ & $-$ & COVID-19 pandemic \\
\bottomrule
\end{tabular}

\smallskip
{\footnotesize Note: no quarter of 2008--2009 appears under any threshold.
Arrival rate $\hat\mu=\PoiMu$ (4 events in 78 quarters); dispersion of
magnitudes $\hat\sigma_J=7.72$ pp, dominated by 2020Q2. Source: own
elaboration.}
\end{table}

\begin{figure}[htbp]
\centering
\figura{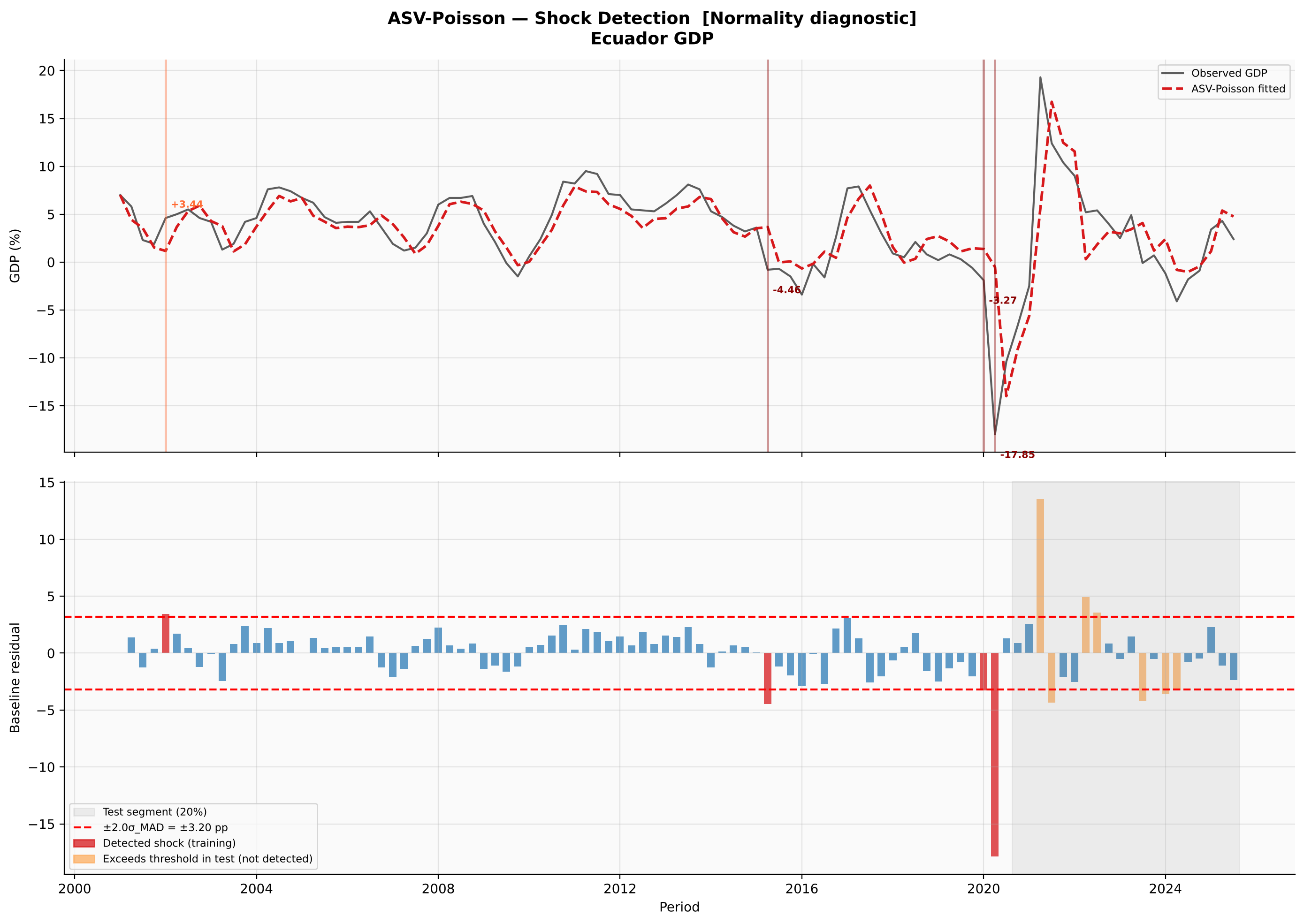}
\caption{Structural shocks detected by the ASV-Poisson. Lower panel: base
ASV residuals, active threshold $\pm2\sigma_{\mathrm{MAD}}=\pm3.20$ pp,
training detections (red) and test-segment exceedances not detected by
design (orange, shaded area).}
\label{fig:shocks}
\end{figure}

The mechanism identifies 2 of the 3 documented episodes (the 2015--2016
slump and the pandemic, the latter in two consecutive quarters), with an
additional positive detection (2002Q1), most plausibly the investment boom
associated with the construction of the Heavy Crude Oil Pipeline
(2001--2003). \textbf{H3a is rejected in its literal formulation.} The
omission of 2008--2009 is, however, a robust result with economic content:
it is invariant to all thresholds ---it does not appear even among the 11
detections of the most permissive one--- because inertia, the annual delay
and the memory kernel absorb that slowdown as regular dynamics. For
Ecuadorian GDP, the global financial crisis was an explainable slowdown
rather than an anomalous break, in contrast with the pandemic, whose
residual is five times the threshold (Figure~\ref{fig:shocks}). The
success is, moreover, retrospective: it establishes internal consistency,
not the ability to anticipate. By design, detection is not run on the test
segment, where 7 of the 20 observations exceed the threshold, including
the 2021Q2 rebound; the catalogue is therefore limited to the training
segment.

The parameters of the jump kernel, $\lambda_J=\PoiLamJ$ and $k_J=\PoiKJ$,
imply $\tau^*_J=(3.730-1)/2.180=\TauJ$ quarters: discrete shocks
materialise quickly, in contrast with the diffuse propagation of the memory
kernel, whose peak lies between 3 and 4 quarters
(Section~\ref{sec:robustez}). The coexistence of a fast impact scale and a
slower propagation scale is a descriptive finding of this work, although
$\tau^*_J$ is estimated from four events dominated by the pandemic and
should be read with caution.

As for residual normality, the ASV-Poisson rejects Shapiro-Wilk with
$p<0.001$, as do the other three models. In the test segment, the
$p$-values are 0.0036 (base ASV) and 0.0052 (ASV-Poisson): a numerical
improvement that does not avoid rejection at 1\,\%. \textbf{H3b is
rejected.}

\subsection{Integrated comparison}
\label{sec:comparacion}

Table~\ref{tab:comparacion} brings the four models together on the
harmonised basis. The SARIMA dominates efficiency (with
$\Delta\mathrm{AIC}=40.13$ against the best ASV), squared and absolute
accuracy, goodness of fit and residual independence, and it does so in all
three volatility regimes, which confirms H1. Among the ASV models there is
no pattern favouring the jump mechanism: the ASV-GP has the lowest error in
P1, the ASV-Poisson in P2 and the deterministic ASV in P3, precisely the
regime of extreme events where the Poisson component should have helped.
No model passes the normality test.

\begin{table}[htbp]
\centering
\caption{Integrated comparison of the four models ($N_{\mathrm{eval}}=\Neval$; harmonised Gaussian basis).}
\label{tab:comparacion}
\small
\begin{tabular}{lccccc}
\toprule
Metric & SARIMA & Det.\ ASV & ASV-GP & ASV-Poisson & Best \\
\midrule
\multicolumn{6}{l}{\textit{Statistical efficiency}} \\
AIC (harm.) & \textbf{\SarAIC} & \AsvAIC & \GpAIC & \PoiAIC & SARIMA \\
BIC (harm.) & \textbf{\SarBIC} & \AsvBIC & \GpBIC & \PoiBIC & SARIMA \\
Parameters ($k$) & \SarK & \AsvK & \GpK & \PoiK & --- \\
\midrule
\multicolumn{6}{l}{\textit{In-sample accuracy}} \\
RMSE (pp) & \textbf{\SarRMSE} & \AsvRMSE & \GpRMSE & \PoiRMSE & SARIMA \\
MAE (pp)  & \textbf{\SarMAE}  & \AsvMAE  & \GpMAE  & \PoiMAE  & SARIMA \\
$R^2$     & \textbf{\SarRdos} & \AsvRdos & \GpRdos & \PoiRdos & SARIMA \\
RMSE P1: 2001--2008 & \textbf{1.281} & 1.399 & 1.363 & 1.387 & SARIMA \\
RMSE P2: 2009--2019 & \textbf{1.439} & 1.729 & 1.728 & 1.709 & SARIMA \\
RMSE P3: 2020--2025 & \textbf{4.479} & 5.297 & 5.372 & 5.338 & SARIMA \\
\midrule
\multicolumn{6}{l}{\textit{Residual diagnostics}} \\
Ljung-Box $p$     & \textbf{\SarLB} & \AsvLB & \GpLB & \PoiLB & SARIMA \\
Shapiro-Wilk $p$  & $<0.001$ & $<0.001$ & $<0.001$ & $<0.001$ & None \\
Excess kurtosis   & \SarKurt & 17.23 & 17.40 & \textbf{16.09} & ASV-Poisson \\
\midrule
\multicolumn{6}{l}{\textit{Structural capabilities}} \\
Out-of-sample 90\,\% coverage & --- & --- & \textbf{\GpCob\,\%} & --- & ASV-GP \\
Shock catalogue & No & No & No & \textbf{Yes} & ASV-Poisson \\
$\tau^*$ (quarters) & --- & \TauEst & --- & \TauJ$^{\dagger}$ & --- \\
\midrule
\multicolumn{6}{l}{\textit{Out-of-sample validation (80/20)}} \\
RMSE (pp) & --- & \textbf{\OosAsv} & --- & \OosPoi & Det.\ ASV \\
\bottomrule
\end{tabular}

\smallskip
{\footnotesize Note: $^{\dagger}$~for the ASV-Poisson we report $\tau^*_J$,
the peak of the jump kernel, which is conceptually different from the
$\tau^*$ of the memory kernel. For the ASV-GP, $\tau^*$ is not defined.
SARIMA coverage was not evaluated with the same protocol (see
Section~\ref{sec:conclusiones}). The SARIMA $k$ includes the variance
$\sigma^2_\varepsilon$ and the ASV $k$ does not; counting it homogeneously
would add 2 units to the AIC of each ASV model, without changing the
ranking. Source: own elaboration.}
\end{table}

In out-of-sample validation, the ASV-Poisson obtains an RMSE of \OosPoi{}
pp against \OosAsv{} for the base ASV: a deterioration of 0.074 pp despite
using two additional parameters. The jump mechanism does not improve
predictive generalisation.

\subsection{Forecasts 2025Q4 to 2027Q3}
\label{sec:pronostico}

\begin{figure}[htbp]
\centering
\figura{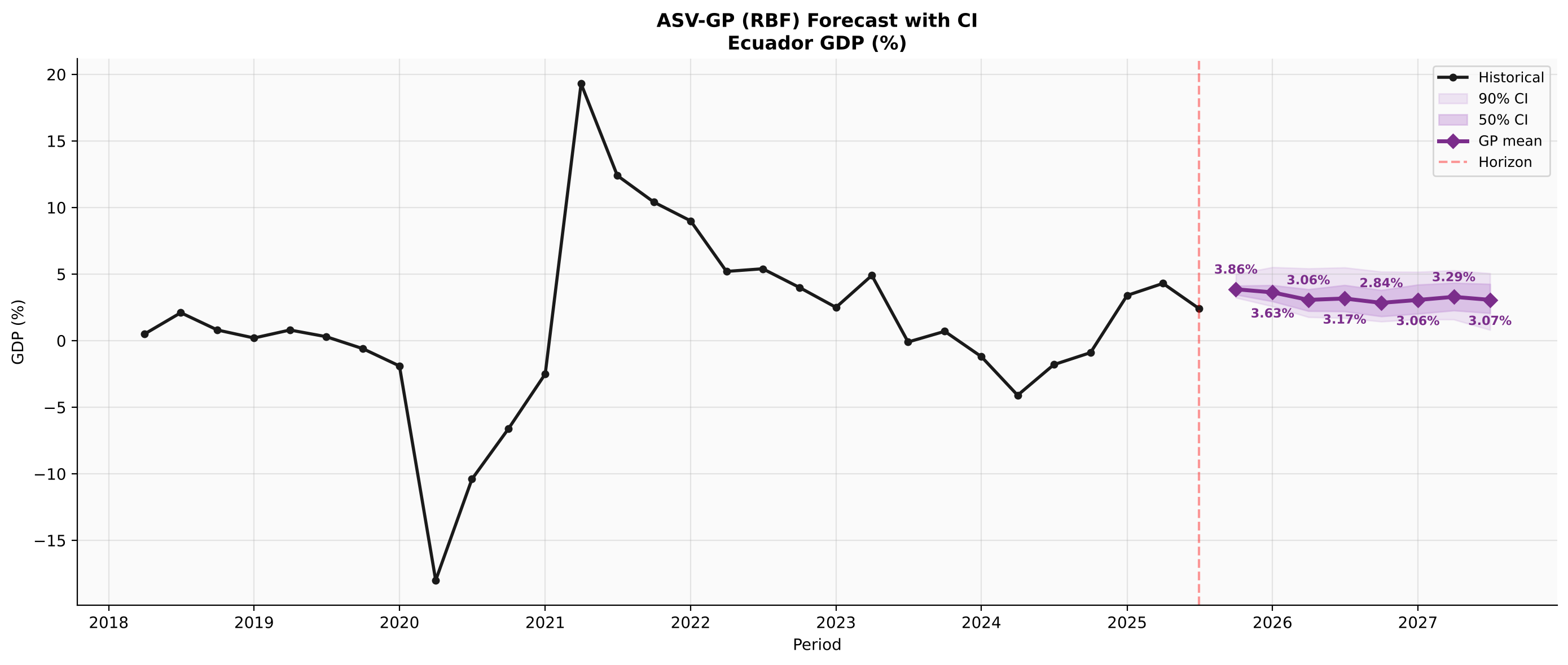}
\caption{ASV-GP forecast with 50\,\% and 90\,\% credible intervals.}
\label{fig:gppron}
\end{figure}

The three ASV models project moderate growth, between 2.2 and 3.9\,\%,
stabilising towards 2027; the SARIMA projects a slowdown with a negative
reading in 2026Q2 (Table~\ref{tab:pronostico}, Figure~\ref{fig:gppron}).
The divergence has an identifiable specification cause: without a
constant, the unconditional SARIMA projection reverts towards zero at rate
$\phi_1^h$, rather than towards the historical mean, and the seasonal
structure at the invertibility boundary imprints the initial oscillating
pattern. Its 90\,\% intervals widen from 8.1 pp in the first quarter to
15.9 pp in 2027Q3, against 4.2 pp for the ASV-GP 90\,\% interval in the same
quarter. That narrower width must be read with the caveat that the
multi-step ASV-GP intervals propagate only the parametric uncertainty of
the kernel, whereas the validated calibration refers to one-step intervals
that add residual noise.

\begin{table}[htbp]
\centering
\caption{Projections of year-on-year real GDP growth (\%).}
\label{tab:pronostico}
\small
\begin{tabular}{lcccc}
\toprule
Period & SARIMA & Det.\ ASV & ASV-GP (mean) & ASV-GP 90\,\% interval \\
\midrule
2025Q4 & 1.72    & 3.66 & 3.86 & $[3.23,\,4.99]$ \\
2026Q1 & $-0.98$ & 2.97 & 3.63 & $[2.58,\,5.51]$ \\
2026Q2 & $-3.95$ & 2.22 & 3.06 & $[1.77,\,5.42]$ \\
2026Q3 & 1.12    & 2.55 & 3.17 & $[1.70,\,5.49]$ \\
2026Q4 & 1.17    & 2.24 & 2.84 & $[1.44,\,5.18]$ \\
2027Q1 & 1.42    & 2.73 & 3.06 & $[1.61,\,5.16]$ \\
2027Q2 & 1.69    & 3.31 & 3.29 & $[1.59,\,5.28]$ \\
2027Q3 & 1.16    & 3.16 & 3.07 & $[0.81,\,5.06]$ \\
\bottomrule
\end{tabular}

\smallskip
{\footnotesize Note: 90\,\% interval = 5th--95th percentiles of 300 GP
posterior trajectories (fixed seed). The ASV-Poisson projects medians very
close to those of the deterministic ASV. Source: own elaboration.}
\end{table}

\subsection{Structural impulse responses}
\label{sec:sirf}

\begin{figure}[htbp]
\centering
\figura{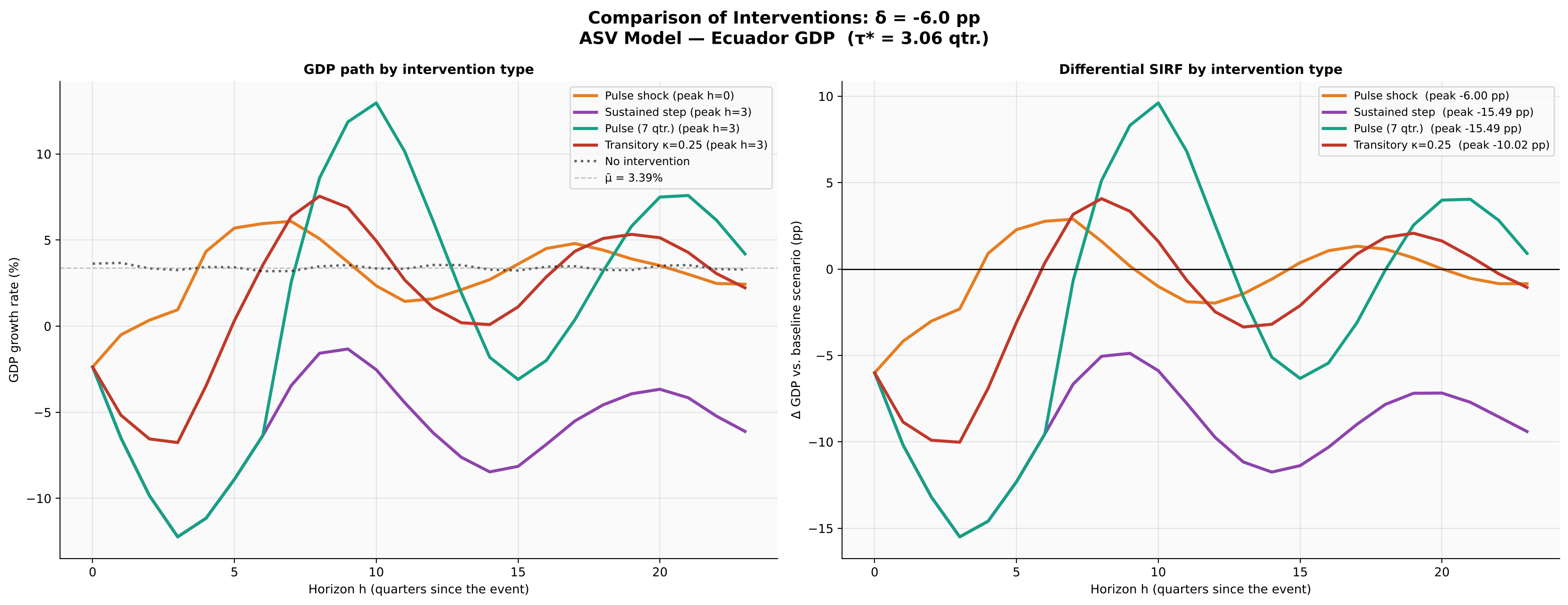}
\caption{Structural impulse response functions of the ASV for an
intervention of $\delta=-6$ pp: one-off shock, sustained intervention,
seven-quarter pulse and transitory shock ($\kappa=0.25$). Left: growth
paths; right: difference from the baseline scenario.}
\label{fig:sirf}
\end{figure}

We constructed a structural impulse response function (SIRF) of the
deterministic ASV by adding an intervention term $\delta\cdot I(t)$ with a
common magnitude $\delta=-6$ pp for four types (Figure~\ref{fig:sirf}).
Because the model is linear in $y$, the responses scale proportionally with
$\delta$.
\begin{itemize}
\item \textbf{One-off shock}: maximum impact at $h=0$ ($-6.00$ pp), effect
halved at $h=3$, sign change at $h=4$ and an endogenous rebound of about
$+2.9$ pp at $h=7$, generated by the annual delay ($\gamma<0$) without
exogenous stimulus. The oscillations dampen slowly and do not fall
permanently below 0.1 pp within the simulated horizon of 24 quarters.
\item \textbf{Sustained intervention}: amplification with a peak of
$-15.49$ pp at $h=3$, a multiplier of 2.58 on $\delta$. The peak coincides
with $\tau^*$, as expected since the SIRF is derived from the same
parameters; it is therefore not an independent validation of the kernel.
By linearity, a favourable intervention of $+3.5$ pp, analogous to a regime
change such as dollarisation, would produce a peak of $+9.04$ pp. After the
peak, the response oscillates between $-5$ and $-12$ pp without settling
within the horizon.
\item \textbf{Seven-quarter pulse} ($h=0$ to $6$): it coincides with the
sustained intervention while active; when withdrawn it generates an
overshoot of about $+9.6$ pp at $h=10$ and damped oscillations with a
period close to 11 quarters.
\item \textbf{Transitory shock} (decay $\kappa=0.25$): transmission lag,
with a peak of $-10.02$ pp at $h=3$.
\end{itemize}
Overall, the estimated dynamics are stable but weakly damped: the negative
annual delay turns any persistent disturbance into multi-year cycles.

\subsection{Robustness: fractional memory}
\label{sec:fraccionario}

Three variants with kernels based on the Riemann-Liouville fractional
derivative were estimated. The best one (pure RL) obtains AIC $=505.17$
($k=7$) and RMSE $=2.966$ pp, without improving on the ASV with a Gamma
kernel. The fractional order converges to the upper bound of the search
space ($d=0.99$), which in this normalised parametrisation corresponds to
the fastest decay in the family, with weights proportional to $1/\tau$:
the series does not demand more memory than the Gamma kernel provides. Its
residuals keep linear dependence under control (Ljung-Box $p=0.081$), but
show an excess kurtosis of 17.76, higher than the SARIMA's: making memory
more flexible, whether Gamma or fractional, does not by itself resolve the
leptokurtosis.

\subsection{Robustness of the structural results}
\label{sec:robustez}

Three additional tests assess whether the results of the ASV framework
depend on estimation choices (Table~\ref{tab:robustez}).

\paragraph{Strict out-of-sample validation.} In the baseline protocol
(Section~4.4) only $\hat\sigma_\varepsilon$, shock detection and
$\lambda_J$, $k_J$ use the training segment exclusively; the structural
parameters, the GP hyperparameters and the mean $\bar\mu$ are estimated
with the full sample. When all of them are re-estimated with $t<79$ only
(2001Q1 to 2020Q3), the ASV-GP 90\,\% interval keeps its 18/20 coverage,
so H2 holds; the 50\,\% interval drops to 7/20 (35\,\%), a value that the
binomial test does not reject ($p=0.26$) but that no longer matches the
nominal level. The test RMSE of the base ASV rises from \OosAsv{} to 4.775
pp, and that of the ASV-Poisson is practically the same (4.773 pp): the
jump mechanism still adds no predictive ability. The shock catalogue
changes (2015Q2, 2017Q1, 2020Q2 and 2020Q3), but no quarter of 2008--2009
appears under this estimation either.

\paragraph{Kernel stability with expanding samples.} With data up to
2020Q3, the estimation sets the memory term to zero ($\beta=0$).
Continuous memory appears as soon as the sample includes the 2021 rebound,
and from then on $\tau^*$ is stable, between 3.0 and 3.3 quarters
(panel B). This explains the deterioration of the base ASV in the strict
test and qualifies the interpretation of the kernel: the estimated
propagation scale rests to a large extent on the 2020--2021 fall and
recovery.

\paragraph{Sensitivity to the search bound for $k$.} The value $k=4$
coincides with the upper bound of the search space. When the bound is
widened, $k$ and $\lambda$ grow together, the AIC improves slightly and
$\tau^*$ rises to 3.76 and 3.89 quarters (panel C). The data therefore
identify the position of the peak (between 3 and 4 quarters), but not the
width of the kernel. The AIC improvement (about 2.5 units) does not alter
the ranking against the SARIMA.

\begin{table}[htbp]
\centering
\caption{Robustness tests of the structural results.}
\label{tab:robustez}
\small
\begin{tabular}{lcccc}
\toprule
\multicolumn{5}{l}{\textit{Panel A. Strict out-of-sample validation (test: 2020Q4 to 2025Q3, $n=20$)}} \\
\midrule
\multicolumn{3}{l}{Metric} & Baseline protocol & Strict re-estimation \\
\midrule
\multicolumn{3}{l}{90\,\% interval coverage (ASV-GP)} & 18/20     & 18/20 \\
\multicolumn{3}{l}{50\,\% interval coverage (ASV-GP)} & 10/20     & 7/20 \\
\multicolumn{3}{l}{Test RMSE, base ASV (pp)}          & \OosAsv   & 4.775 \\
\multicolumn{3}{l}{Test RMSE, ASV-Poisson (pp)}       & \OosPoi   & 4.773 \\
\midrule
\multicolumn{5}{l}{\textit{Panel B. Deterministic ASV with expanding samples}} \\
\midrule
End of sample & $n$ & $\beta$ & $k$ & $\tau^*$ \\
\midrule
2020Q3 & 79 & 0.000 & 1.53 & --- \\
2021Q3 & 83 & 0.114 & 4.00$^{\dagger}$ & 3.33 \\
2022Q3 & 87 & 0.064 & 4.00$^{\dagger}$ & 3.22 \\
2023Q3 & 91 & 0.074 & 4.00$^{\dagger}$ & 3.11 \\
2024Q3 & 95 & 0.078 & 4.00$^{\dagger}$ & 3.04 \\
2025Q3 & 99 & 0.077 & 4.00$^{\dagger}$ & \TauEst \\
\midrule
\multicolumn{5}{l}{\textit{Panel C. Sensitivity to the upper bound for $k$ (full sample)}} \\
\midrule
Bound for $k$ & $k$ & $\lambda$ & $\tau^*$ & AIC (harm.) \\
\midrule
$\leq 4$ (baseline) & 4.00$^{\dagger}$ & 0.98 & \TauEst & \AsvAIC \\
$\leq 8$            & 8.00$^{\dagger}$ & 1.86 & 3.76 & 502.16 \\
$\leq 15$           & 8.78 & 2.00$^{\dagger}$ & 3.89 & 501.84 \\
\bottomrule
\end{tabular}

\smallskip
{\footnotesize Note: $^{\dagger}$~value at the bound of the search space.
In panel B, with $\beta=0$ the memory term vanishes and $\tau^*$ is not
defined. In the strict re-estimation, all parameters (including $\bar\mu$)
are estimated with $t<79$. Source: own elaboration.}
\end{table}

\FloatBarrier
\subsection{Hypothesis tests}

Table~\ref{tab:hipotesis} summarises the test of the three hypotheses in
light of the previous results, including the robustness tests.

\begin{table}[htbp]
\centering
\caption{Outcome of the hypothesis tests.}
\label{tab:hipotesis}
\small
\begin{tabular}{clp{7.2cm}l}
\toprule
Hyp. & Statement & Evidence & Outcome \\
\midrule
H1  & SARIMA outperforms ASV & AIC \SarAIC{} vs.\ \AsvAIC; RMSE \SarRMSE{} vs.\ \AsvRMSE; $R^2$ \SarRdos{} vs.\ \AsvRdos & Confirmed \\
H2  & ASV-GP 90\,\% interval calibrated & 90\,\%: 18/20, also under strict re-estimation; 50\,\%: 10/20 (baseline) and 7/20 (strict) & Confirmed \\
H3a & Detection of the 3 episodes & 2/3; robust omission of 2008--2009; extra detection in 2002Q1 & Rejected \\
H3b & Improved residual normality & Shapiro-Wilk $p<0.001$ in all four models; no out-of-sample improvement & Rejected \\
\bottomrule
\end{tabular}
\end{table}
\FloatBarrier

% ====================================================================
\section{Results with shock covariates}
\label{sec:covariables}
% ====================================================================

\subsection{Estimation and selection}

Table~\ref{tab:cov_ajuste} compares the specifications on the same
harmonised basis ($N_{\mathrm{eval}}=\Neval$). The stringency index is, on
its own, the decisive covariate: it reduces the SARIMA AIC by 57.6 units,
lowers residual kurtosis from 16.6 to 1.1, and the Shapiro-Wilk test no
longer rejects normality at 5\,\%. Adding oil improves the AIC by a further
3.8 units, and that is the specification selected. In it,
$\hat\beta_{\mathrm{COVID}}=-19.04$ (SE 2.06) and
$\hat\beta_{\mathrm{WTI}}=3.12$ (SE 0.92), both with $p<0.001$: the 87-point
rise of the index in 2020Q2 is equivalent to about $-16.6$ pp of growth,
and a 10\,\% year-on-year rise in the oil price to about $+0.30$ pp. The
financial covariate is not significant ($\hat\beta=-2.36$, SE 3.41,
$p=0.49$) and worsens the AIC: from this perspective too, the 2008--2009
crisis does not appear as a specific shock to Ecuadorian GDP, consistent
with the findings of Section~\ref{sec:poisson}.

Within the Volterra framework, the ASV-X improves the ASV AIC by 39.7 units
and reduces its residual kurtosis from 17.2 to 2.1, although without
reaching the efficiency of the SARIMAX. Its propagation peak
($\tau^*=3.43$ quarters) lies within the 3-to-4-quarter range of
Section~\ref{sec:robustez}. Its residuals, however, remain autocorrelated
(Ljung-Box $p=0.003$): the internal dynamics of the ASV propagate part of
the covariates' effect, which suggests an interaction between memory and
covariate that the contemporaneous form does not capture.

\begin{table}[htbp]
\centering
\caption{Models with shock covariates (harmonised basis, $N_{\mathrm{eval}}=\Neval$).}
\label{tab:cov_ajuste}
\small
\begin{tabular}{lcccccc}
\toprule
Model & $k$ & AIC & RMSE (pp) & Ljung-Box $p$ & Shapiro-Wilk $p$ & Kurtosis \\
\midrule
SARIMA (no covariates)                   & 4  & \SarAIC & \SarRMSE & \SarLB & $<0.001$ & \SarKurt \\
SARIMAX: COVID                           & 5  & 406.68 & 1.831 & 0.264 & 0.058 & 1.13 \\
\textbf{SARIMAX: COVID + oil}            & 6  & \textbf{402.86} & \textbf{1.778} & 0.535 & 0.124 & 1.26 \\
SARIMAX: COVID + oil + finance           & 7  & 405.28 & 1.781 & 0.569 & 0.087 & 1.36 \\
\midrule
ASV (no covariates)                      & 8  & \AsvAIC & \AsvRMSE & \AsvLB & $<0.001$ & 17.23 \\
ASV-X: COVID + oil                       & 10 & 464.66 & 2.339 & 0.003 & 0.026 & 2.12 \\
\bottomrule
\end{tabular}

\smallskip
{\footnotesize Note: SARIMAX with the $(1,0,0)(1,0,1)_4$ order of the
reference model; $k$ includes the variance in the SARIMAX and the intercept
in the ASV-X. In bold, the specification selected by AIC. Source: own
elaboration.}
\end{table}

\subsection{Shock attribution and residual normality}

\begin{figure}[htbp]
\centering
\figura{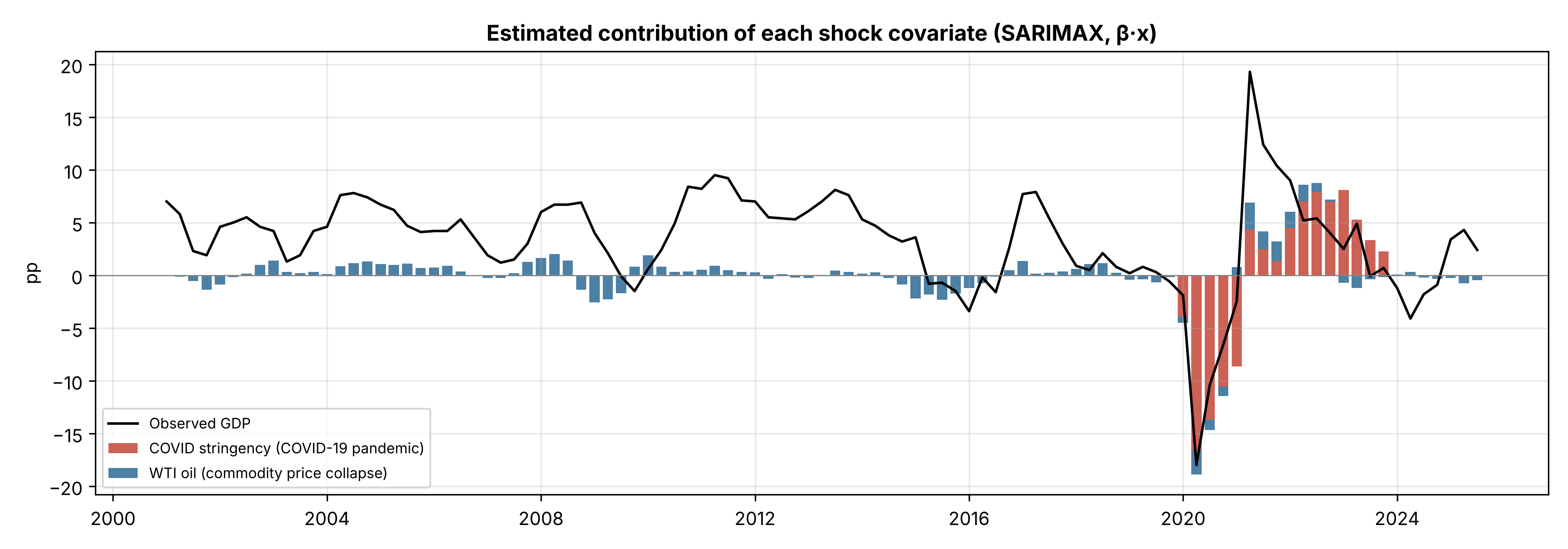}
\caption{Estimated contribution of each shock covariate to GDP growth
(SARIMAX, $\hat\beta_m x_{m,t}$) and observed growth.}
\label{fig:contribuciones}
\end{figure}

\begin{figure}[htbp]
\centering
\figura{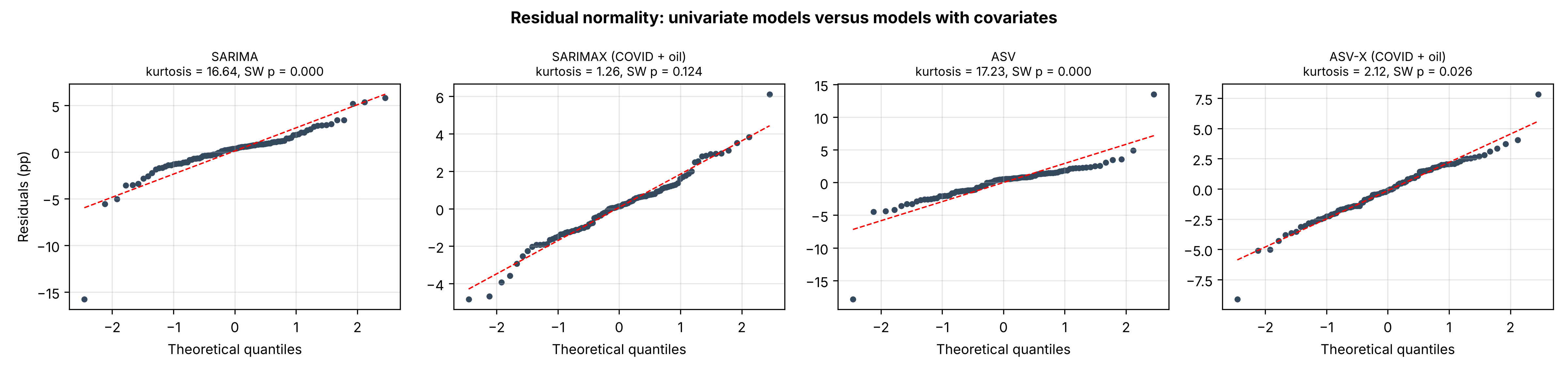}
\caption{Q-Q plots of the residuals of the univariate models and of their
versions with shock covariates.}
\label{fig:normalidad}
\end{figure}

Figure~\ref{fig:contribuciones} decomposes the estimated contribution of
each covariate ($\hat\beta_m x_{m,t}$). Stringency explains the 2020Q2
collapse ($-16.6$ pp out of an observed total of $-18.0$) and contributes
positively during the recovery; within the pandemic window its mean
contribution is $-5.7$ pp. Oil explains a smaller part of the 2015--2016
slump (mean contribution of $-1.2$ pp and minimum of $-2.3$ pp), which
indicates that this episode had domestic components that the external
covariate does not capture. With the robust threshold of
Section~\ref{sec:poisson}, 2020Q2 is no longer among the extreme training
residuals, while 2015Q2 remains and 2008Q4 and 2011Q2 appear.
Figure~\ref{fig:normalidad} shows the most relevant result: with
covariates, the residual tails line up with the normal distribution,
something that neither the univariate models nor the jump component
achieved.

\subsection{Out-of-sample performance}

With the strict split (everything estimated with $t<79$), the SARIMAX
reduces the test RMSE from 3.247 to 3.154 pp, whereas the ASV-X raises it
from 4.590 to 4.798 pp (Table~\ref{tab:cov_oos}). The reason is the same as
in Section~\ref{sec:robustez}: the training segment ends in 2020Q3 and
contains only three quarters with stringency, so its effect is poorly
identified. With the rolling origin, which re-estimates the model at each
quarter of the test segment and uses the observed covariates, the SARIMAX
reduces the SARIMA error at all four horizons, by between 11 and 15\,\%.

\begin{table}[htbp]
\centering
\caption{Out-of-sample performance with shock covariates (RMSE, pp).}
\label{tab:cov_oos}
\small
\begin{tabular}{lcccccc}
\toprule
 & \multicolumn{2}{c}{Strict split} & \multicolumn{4}{c}{Rolling origin (SARIMA)} \\
\cmidrule(lr){2-3}\cmidrule(lr){4-7}
Model & No cov. & With cov. & $h=1$ & $h=2$ & $h=3$ & $h=4$ \\
\midrule
SARIMA / SARIMAX & 3.247 & 3.154 & & & & \\
\quad No covariates   & & & 3.317 & 4.577 & 5.602 & 5.899 \\
\quad With covariates & & & 2.955 & 3.915 & 4.935 & 5.140 \\
ASV / ASV-X      & 4.590 & 4.798 & --- & --- & --- & --- \\
\bottomrule
\end{tabular}

\smallskip
{\footnotesize Note: strict split with all parameters estimated on $t<79$;
expanding-window rolling origin from $t=79$ with covariates observed over
the horizon. The rolling origin of the ASV-X was not computed because of
its computational cost. Source: own elaboration.}
\end{table}

\subsection{Ex post evaluation and updated forecast}

After the sample closed, the Central Bank of Ecuador published growth of
5.0\,\% for 2025Q4, 2.1\,\% for 2026Q1 and 2.1\,\% for 2026Q2
\citep{bce2026}. Table~\ref{tab:expost} compares these values with each
model's forecasts from 2025Q3, conditional on the covariates observed over
the horizon. The result reverses the in-sample ranking: the univariate ASV
models are the most accurate (MAE of 0.77 to 0.78 pp), followed by the
ASV-GP (1.21) and the ASV-X (1.30), whereas the SARIMA (4.14) and the
SARIMAX (3.30) forecast a contraction that did not occur, owing to the
reversion towards zero inherent in their specification without a constant
(Section~\ref{sec:pronostico}). The covariates improve the SARIMA but do
not correct that bias; in the ASV-X they worsen the forecast because the
2026 oil price increase (+48\,\% year on year in 2026Q2) did not translate
into higher growth. No model anticipates the 5.0\,\% of 2025Q4, associated
with an export upturn. With only three preliminary observations, the
evaluation is indicative.

\begin{table}[htbp]
\centering
\caption{Ex post evaluation against published GDP (origin 2025Q3; \%).}
\label{tab:expost}
\small
\begin{tabular}{lccccccc}
\toprule
Quarter & Actual & SARIMA & SARIMAX & ASV & ASV-X & ASV-GP & ASV-Poisson \\
\midrule
2025Q4 & 5.0 & 1.72    & 0.93    & 3.66 & 2.94 & 3.86 & 3.66 \\
2026Q1 & 2.1 & $-0.98$ & $-1.03$ & 2.97 & 2.52 & 3.63 & 2.96 \\
2026Q2 & 2.1 & $-3.95$ & $-0.60$ & 2.22 & 3.51 & 3.06 & 2.21 \\
\midrule
MAE (pp)  & & 4.14 & 3.30 & 0.78 & 1.30 & 1.21 & \textbf{0.77} \\
RMSE (pp) & & 4.35 & 3.35 & 0.93 & 1.46 & 1.23 & \textbf{0.92} \\
\bottomrule
\end{tabular}

\smallskip
{\footnotesize Note: SARIMAX and ASV-X use the observed WTI values over the
horizon and zero stringency. ASV-GP and ASV-Poisson correspond to
Table~\ref{tab:pronostico}. Preliminary actual data \citep{bce2026}.
Source: own elaboration.}
\end{table}

\begin{figure}[htbp]
\centering
\figura{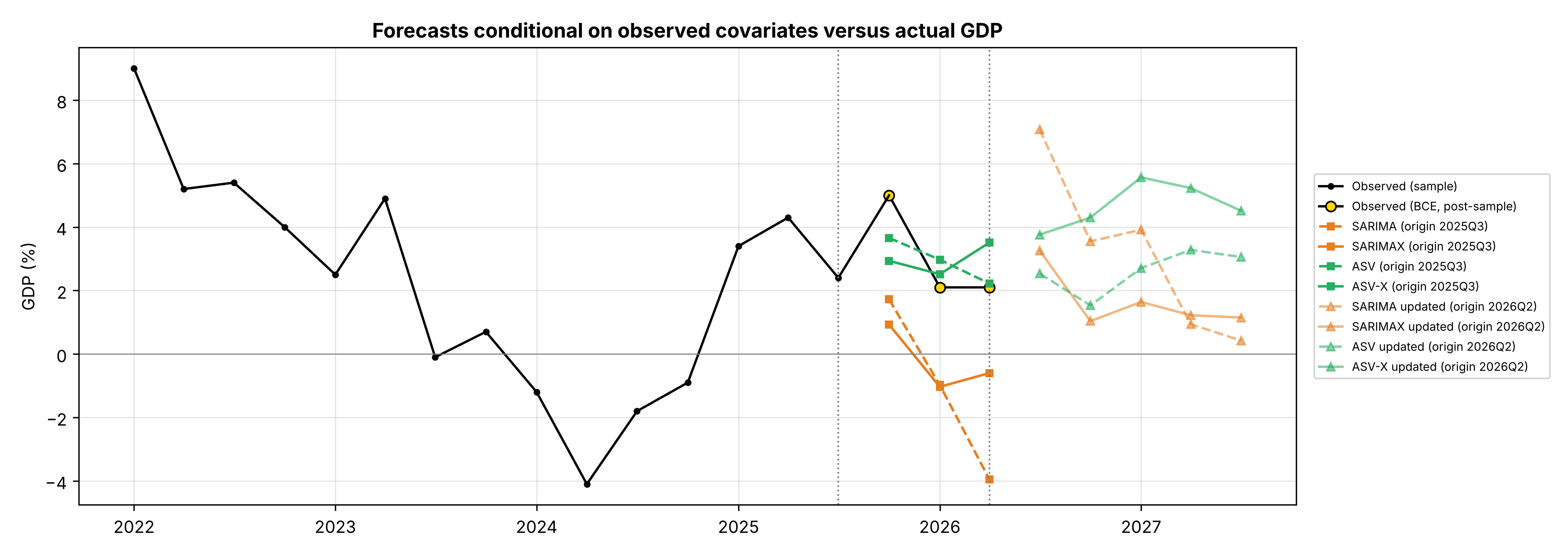}
\caption{Forecasts from 2025Q3 conditional on the observed covariates, GDP
published by the Central Bank of Ecuador, and updated forecast from
2026Q2.}
\label{fig:expost}
\end{figure}

Finally, the models are updated with the three published data points,
without re-estimating their parameters, and the forecast is made from
2026Q2 (Table~\ref{tab:actualizado}, Figure~\ref{fig:expost}). For 2026Q3
the forecast is conditional on the oil price actually observed in that
quarter (+32\,\% year on year). The ASV models project growth of 2.5 to
3.8\,\% for that quarter, whereas the univariate SARIMA jumps to 7.1\,\%
because of the instability of its seasonal component at the invertibility
boundary.

\begin{table}[htbp]
\centering
\caption{Updated forecast from 2026Q2 (\%).}
\label{tab:actualizado}
\small
\begin{tabular}{lcccc}
\toprule
Quarter & SARIMA & SARIMAX & ASV & ASV-X \\
\midrule
2026Q3 & 7.08 & 3.25 & 2.54 & 3.75 \\
2026Q4 & 3.55 & 0.60 & 1.53 & 3.78 \\
2027Q1 & 3.91 & 1.20 & 2.71 & 4.80 \\
2027Q2 & 0.94 & 0.79 & 3.28 & 4.31 \\
2027Q3 & 0.42 & 0.72 & 3.06 & 3.48 \\
\bottomrule
\end{tabular}

\smallskip
{\footnotesize Note: parameters estimated with 2001Q1 to 2025Q3; state
updated with the data for 2025Q4 to 2026Q2. Covariates: WTI observed up to
2026Q3 and constant afterwards; zero stringency. Source: own elaboration.}
\end{table}

\FloatBarrier

% ====================================================================
\section{Discussion}
\label{sec:discusion}
% ====================================================================

\paragraph{Scope of the SARIMA's superiority.} The SARIMA's advantage is
the most robust result of the study: it holds for all point metrics, in all
three regimes, and regardless of whether errors are penalised
quadratically or linearly. It is consistent with the evidence that
well-specified SARIMA models outperform more complex alternatives in point
accuracy when short-run memory and seasonality are stable
\citep{petropoulos2022}. Its practical magnitude, however, is moderate:
0.442 pp of RMSE (17.8\,\%) on a series with a range above 37 pp, a
difference that would rarely change a policy decision with error margins
of one to two points.

\paragraph{Residual kurtosis as a shared limitation.} The best-fitting
model is also the one that least satisfies the distributional assumption
of its own intervals: without a jump mechanism, the pandemic quarters
remain as extreme residuals. This is precisely the problem that motivates
the intervention analysis of \citet{boxtiao1975}: an ARIMA without
explanatory variables treats discrete events as noise. Even so, the
evidence does not allow this argument to be turned into a victory for the
ASV framework: no model passes the normality test, and the fractional
variant has an even higher kurtosis. No univariate specification
internalises the non-normality; Section~\ref{sec:covariables} shows that a
measured covariate does: the excess kurtosis came from an identifiable
shock, not from an irreducible property of the process.

\paragraph{Calibration of the ASV-GP.} The result shows that it is
possible to propagate uncertainty about the memory weights and obtain
intervals that reflect the probability of error in a validation segment
that includes the post-pandemic rebound, even when all parameters are
re-estimated with the training segment only. That RBF and Matérn 5/2 are
indistinguishable suggests that the memory weights vary in an effectively
smooth way, with no local features that the roughness of the Matérn could
exploit \citep{rasmussen2006}. The ASV-GP does not buy point accuracy, but
uncertainty quantification.

\paragraph{The Poisson component as a diagnostic tool.} The three
unfavourable results ---non-exhaustive detection, no improvement in
normality and worse out-of-sample performance--- converge: the Poisson
component is a retrospective diagnostic tool, not a mechanism for
improving the model. Its legitimate contribution is the parametric
characterisation of the detected breaks and the evidence, robust to every
threshold, that the 2008--2009 crisis was absorbed by the regular dynamics
of Ecuadorian GDP. The function expected from the jump component
---absorbing the pandemic collapse and normalising the residuals--- is
fulfilled by the stringency covariate, with a single interpretable
parameter.

\paragraph{Scope of the memory scale.} The propagation scale is the most
exposed structural result. Its position, between 3 and 4 quarters,
withstands the widening of the search bound for $k$ and the extension of
the sample from 2021 onwards; its point value and width do not, and memory
is identified only when the sample includes the 2020--2021 cycle
(Section~\ref{sec:robustez}). The interpretation in terms of dollarisation
is therefore a hypothesis compatible with the data rather than an
established finding: testing it requires samples not dominated by a single
episode, or comparison with similar economies.

\paragraph{Explaining versus forecasting.} Shock covariates and continuous
memory solve different problems. The former explain the past: they
attribute the 2020 collapse to stringency, a fraction of the 2015--2016
slump to oil, and rule out an external financial factor in 2008--2009. But
their usefulness for forecasting depends on the factor that originated the
shock moving again, and doing so with the same effect: in 2026 the oil
price rose sharply without growth following it. The continuous memory of
the ASV, anchored by an explicit intercept, was instead what best
forecast the quarters after the sample. The SARIMA's dominance in one-step
metrics did not carry over to multi-step forecasting, not even with
covariates.

\paragraph{Complementarity.} Of the dimensions evaluated, the SARIMA leads
the five point metrics (AIC, RMSE, MAE, $R^2$ and Ljung-Box), the ASV-GP
the calibrated quantification of uncertainty and the ASV-Poisson shock
detection. No model dominates in all of them, which supports the argument
of \citet{hendry2014}: selection cannot rest solely on an information
criterion when the models capture different aspects of the data
generating process. In operational terms, the SARIMA provides the one-step
point forecast; the ASV-GP, calibrated intervals; and the ASV SIRF,
recovery scales (a half-life of 3 quarters, with damped oscillations that
persist for more than five years).

% ====================================================================
\section{Conclusions}
\label{sec:conclusiones}
% ====================================================================

Year-on-year growth of Ecuador's real GDP between 2001 and 2025 is an
$I(0)$ process with short-run inertia, stochastic seasonality and
leptokurtosis concentrated in the pandemic. On a harmonised likelihood
basis, the SARIMA$(1,0,0)(1,0,1)_4$ outperforms the three Volterra models
in all point metrics. The ASV framework contributes, instead, three
elements that the SARIMA does not offer: a shock propagation scale of three
to four quarters, identified mainly from the 2020--2021 cycle, so that its
attribution to real-sector adjustment under dollarisation is tentative;
90\,\% predictive intervals calibrated out of sample, also under strict
re-estimation; and an auditable catalogue of breaks, whose most
informative finding is that the 2008--2009 crisis is not an anomaly for
the internal dynamics of GDP. Modelling shocks as random jumps improved
neither fit nor prediction.

Linking each shock to the covariate that originates it does change the
diagnosis. The stringency index explains the 2020 collapse and, together
with the oil price, reduces the SARIMA AIC to 402.86 and removes the
residual leptokurtosis; the financial indicator is not significant, so the
absence of an anomaly in 2008--2009 is confirmed from another angle.
Against the data published after the sample, however, the univariate ASV
forecast better than every specification with covariates: covariates
explain the shocks that occurred, and continuous memory sustains
multi-step forecasting better.

\paragraph{Limitations.} (i)~With \Nobs{} observations, the GP
hyperparameters and the Poisson process (four events, dominated by 2020)
are estimated with little information, and the binomial test with $n=20$
has limited power. (ii)~The main models are univariate, and the extension
includes only three external covariates, so domestic channels (fiscal,
credit, energy) cannot be disaggregated. (iii)~The empirical-Bayes
approximation fixes the GP hyperparameters and may underestimate total
uncertainty. (iv)~The detector's success is retrospective with respect to
known episodes. (v)~All models assume homoskedasticity, which is violated
across regimes. (vi)~We did not include a SARIMA with heavy-tailed
innovations, nor did we evaluate the coverage of the SARIMA intervals with
the same protocol, which prevents a direct comparison of its calibration
with that of the ASV-GP. (vii)~The estimated Gamma kernel lies at the
search bound for $k$: the position of the peak is identified, but not the
width of the kernel. (viii)~The stringency effect is identified from few
quarters (2020--2022) and assumes a zero index from 2023; the ASV-X retains
residual autocorrelation; and the ex post evaluation uses three
preliminary data points subject to revision.

\paragraph{Future work.} Natural extensions are: full Bayesian inference
of the GP kernel through MCMC; a GARCH component in the error term, since
leptokurtosis might be better represented as conditional variance than as
jumps; a Poisson process with time-varying intensity driven by
vulnerability covariates; widening the shock--covariate register with
Central Bank of Ecuador series (oil production, remittances, electricity
generation for the 2024 energy crisis) and modelling the interaction
between memory and covariates in the ASV-X; prospective evaluation of the
detector on new data; and comparison of $\tau^*$ with other dollarised
economies, in samples not dominated by a single episode.

\section*{Acknowledgements}
The author thanks Dr.\ Christian Eduardo Galarza Morales, supervisor of the
master's thesis from which this paper is derived, for the technical review
that led to the methodological corrections applied here.

\section*{Data and code availability}
GDP data come from the National Accounts Statistics of the Central Bank of
Ecuador \citep{bce2025,bce2026}. The shock covariates come from
\citet{hale2021} (stringency index), \citet{eia2026} (WTI) and
\citet{shiller2015} (S\&P 500); the extension script downloads them from
their public repositories. Analyses were carried out in Python 3.12
(\texttt{statsmodels}, \texttt{scipy}, \texttt{numpy}, \texttt{pandas},
\texttt{matplotlib}) with fixed seeds (Differential Evolution and GP
posterior: 42; interval coverage: 123). The scripts are available from the
author upon request; the robustness tests of Section~\ref{sec:robustez}
are reproduced by enabling the corresponding option in the Volterra suite
script.

% ====================================================================
%  REFERENCES (shared by both language versions)
% ====================================================================

% ====================================================================
\appendix
\section{SARIMA specification and harmonisation}
\label{app:sarima}
% ====================================================================

The reference model, without a constant term, is
\begin{equation}
(1-\phi_1L)(1-\Phi_1L^4)\,y_t=(1+\Theta_1L^4)\,\varepsilon_t,\qquad
\varepsilon_t\sim\mathcal{N}(0,\sigma^2_\varepsilon).
\end{equation}
It was estimated with \texttt{statsmodels} (\texttt{SARIMAX}) by exact
maximum likelihood through the Kalman filter, with the L-BFGS optimiser
(\texttt{maxiter}$\,=200$), without imposing stationarity or invertibility
---which allows $\Theta_1$ to converge to the boundary--- and with
standard errors from the outer product of gradients.

The native AIC (\SarAICnat) is computed on the exact log-likelihood of the
filter with \Nobs{} observations and is not comparable with that of the
ASV models, which is based on one-step prediction residuals. With
Equation~\eqref{eq:aic}, $n=\Neval$ and the constant
$n(1+\ln2\pi)=278.11$, we obtain $\mathrm{AIC}_{\mathrm{harm}}=\SarAIC$ and
$\mathrm{BIC}_{\mathrm{harm}}=\SarBIC$. We take $k=4$, the estimated
parameters ($\phi_1,\Phi_1,\Theta_1,\sigma^2_\varepsilon$); the model has
no constant. The 11.76-unit difference between the two versions reflects
the different likelihood basis, not a change in the model.

On the \Neval{} one-step residuals: Ljung-Box $p=\SarLB$ (10 lags),
Durbin-Watson $=2.098$, MAE $=\SarMAE$ pp, Jarque-Bera $=1253.6$ and
heteroskedasticity test $H=18.27$ ($p<0.01$). The model is a valid linear
filter for the conditional mean, but distributionally misspecified in the
presence of discrete shocks.

\end{document}